# Atomistic understanding of two-dimensional monatomic phase-change material for non-volatile optical applications

Hanyi Zhang[1,#], Xueqi Xing[1,#], Jiang-Jing Wang[1,*], Chao Nie[1], Yuxin Du[1], Junying Zhang[1], Xueyang Shen[1], Wen Zhou[1,*], Matthias Wuttig[2,3], Riccardo Mazzarello[4,*], Wei Zhang[1,*]

[1]Center for Alloy Innovation and Design (CAID), State Key Laboratory for Mechanical Behavior of Materials, Xi'an Jiaotong University, Xi'an, China.

[2]Institute of Physics IA, JARA-FIT, RWTH Aachen University, 52074 Aachen, Germany

[3]Peter Grünberg Institute (PGI 10), Forschungszentrum Jülich GmbH, 52425 Jülich, Germany

[4]Department of Physics, Sapienza University of Rome, Rome 00185, Italy

Emails: j.wang@xjtu.edu.cn, wen.zhou@mail.xjtu.edu.cn, riccardo.mazzarello@uniroma1.it, wzhang0@mail.xjtu.edu.cn

[#]These authors contributed equally to the work.

## Abstract

Elemental antimony (Sb) is a promising material for phase-change memory, neuromorphic computing and nanophotonic applications, because its compositional simplicity can prevent phase segregation upon extensive programming. Scaling down the film thickness is a necessary step to prolong the lifetime of amorphous Sb, but the optical properties of Sb are also significantly altered as the thickness is reduced to a few nanometers, adding complexity to device optimization. In this work, we aim to provide atomistic understanding of the thickness-dependent optical responses in Sb thin films. As thickness decreases, both the extinction coefficient and optical contrast reduce in the near-infrared spectrum, consistent with previous optical measurements. Such thickness dependence gives rise to a bottom thickness limit of 2 nm in photonic applications, as predicted by coarse-grained device simulations. Further bonding analysis reveals a fundamentally different behavior for amorphous and crystalline Sb upon downscaling, resulting in the reduction of optical contrast. Thin film experiments are also carried out to validate our predictions. The thickness-dependent optical trend is fully demonstrated by our ellipsometric spectroscopy experiments, and the bottom thickness limit of 2 nm is confirmed by structural characterization experiments. Finally, we show that the greatly improved amorphous-phase stability of the 2 nm Sb thin film enables robust and reversible optical switching in a silicon-based waveguide device.

## Keywords

phase-change materials, antimony, optical properties, thin films, metavalent bonding

# 1. Introduction

To deal with the drastically increased demands in data storage and processing, massive research efforts have been undertaken to develop non-volatile memory and neuromorphic computing technologies[1-5]. Phase-change materials (PCMs) are a leading material candidate for these applications[3, 6-11]. The basic principle is to exploit the large change in electrical and optical properties associated with the rapid and reversible phase transition between the amorphous and crystalline phases of PCMs. Conventional PCMs contain multiple elements, such as the flagship GeTe-$Sb_2Te_3$ pseudo-binary compounds[12-14], the typical growth-dominant material Ag-In-Sb-Te[15-17], the $Sc_{0.2}Sb_2Te_3$ alloy with ultra-high nucleation rate[18-22], and various new PCMs obtained by computational screening[23-25]. Upon massive programming, phase segregation could occur in these multi-component PCMs, causing device failures[26, 27]. A promising strategy is to use only one element, e.g., Sb, for phase-change applications[28]. Although the crystallization speed of Sb can be exceedingly fast, i.e., Sb-based electronic devices can be crystallized using sub-nanosecond electrical pulse[29], this single-element glass is highly unstable and crystallizes spontaneously at room temperature[30].

By scaling down the film thickness to a few nanometers, the lifetime of amorphous Sb can be prolonged to tens of hours at room temperature attributed to the confinement effects induced by the surrounding layers[28, 31-33]. Such effects were theoretically supported by both *ab initio* molecular dynamics (AIMD) and machine-learning-driven molecular dynamics (MLMD) simulations, providing atomistic understanding of the structural and dynamical properties of nano-confined Sb[28, 34]. It was also demonstrated that few-nm amorphous Sb films can be stabilized in thin film devices and waveguide devices without any capping layer[35-37]. AIMD simulations of amorphous Sb in bulk and thin film forms revealed an increasing structural dissimilarity when approaching the surface regions[38], leading to a crystallization-suppressed slab of ~0.7 nm per surface. This observation indicates that the bulk-like interior could already be smaller than the critical nucleus size[34, 39] in two-dimensional Sb thin films. Therefore, this stabilization mechanism is an intrinsic thickness-dependent effect, regardless the choice of surrounding materials.

In addition to crystallization dynamics, the physical properties of Sb are also largely altered upon thickness downscaling[35], but so far there is no theoretical report on comparing the thickness-dependent optical properties of the amorphous and crystalline phases, which hinders further development of Sb-based photonic devices. Take the PCM-based waveguide device for instance, it turns ON (with high transmission) when the PCM is switched to the amorphous phase, and turns OFF (with low transmission) upon crystallization[40-44]. A functional all-optical waveguide device requires both a sizable optical contrast in the telecom band between the amorphous phase and the crystalline phase and a non-negligible extinct coefficient in both phases to enable sufficient Joule heating for phase transition. However, there should exist a bottom thickness limit for Sb, because as the film thickness is reduced towards the two-dimensional limit, crystalline Sb undergoes a major change in electronic structure from metallic to semiconducting[45], leading to a high transmission state[46] in the monolayer and few-layer Sb[47]. Besides, the optical changes in the visible light range can also be exploited for non-volatile color display applications [35, 48].

In this work, we perform systematic density functional theory (DFT) calculations to understand the thickness dependence of electronic structure and optical profiles in both crystalline (c-) and amorphous (a-) Sb films from sub-nm up to ~5.1 nm, enabling a direct comparison with the spectroscopic ellipsometry measurements. In combination with the coarse-grained finite-element method (FEM) modeling, our multiscale simulations predict a bottom limit of ~2 nm with sufficient programming window for a functional Sb waveguide device. We provide an explanation on the trends of optical variations from a chemical bonding perspective. Moreover, we carry out structural and optical experiments on Sb thin films and confirm the bottom thickness limit of Sb thin films to be ~2 nm, which corresponds to only 12 atomic layers in the crystalline phase, as evidenced by cross-sectional scanning transmission electron microcopy (STEM) experiments. Our electrical and optical measurements prove that this two-dimensional Sb thin film still has sizable contrast in physical properties upon crystallization, which are sufficiently large to enable practical phase-change applications.

## 2. Results and Discussion

It is straightforward to build slab models for c-Sb by adding a vacuum slab of 2.5 nm along the z-axis. We considered a set of models ranging from 1 bilayer (BL) antimonene up to 14 BL slab models (Fig. 1a) with a film thickness of ~0.2–5.1 nm. It is noted that in the case of 1 BL, the structure can be viewed as a buckled monolayer. Hence, antimonene is also termed as a monolayer (ML) in the literature. For 2 BL and thicker models, there is non-negligible covalent interaction between the BLs due to the long Sb–Sb bonds of ~3.35 Å, similarly to the case of GeTe slabs[49]. In the following, we denote the building blocks as "BL" instead of "ML". As regards the amorphous phase, three melt-quenched a-Sb bulk models with independent thermal histories were first generated using AIMD[38]. Each model contained 360 atoms in a box of 2.26×2.18×2.28 nm$^3$. Next, we constructed a-Sb slab models according to the thickness of c-Sb models. For instance, we simulated cutting a slab of ~0.6 nm (size of the 2BL c-Sb model) out of a bulk a-Sb model, and annealed such model in the presence of a 2.5 nm vacuum slab at 300 K over 30 ps. We denoted this a-Sb slab model as "2 BL" as well (Fig. 1b). This a-Sb model of ~0.6 nm basically represents the bottom limit in thickness. The "1 BL" a-Sb model cannot be generated properly, because Sb atoms tend to form cluster laterally (Fig. S1). For thicker models (>2.28 nm), we used two and even three bulk a-Sb models to build the slab. Two or three sub-slabs taken from different bulk models were glued together, and the interfaces between different sub-slabs were properly relaxed to avoid too short chemical bonds or big voids. The largest a-Sb slab model we constructed was "14 BL" (~5.1 nm), containing 798 atoms (Fig. 1b). For each thickness, two additional a-Sb slab models were generated for statistics. For thick models (>2.28 nm), additional models were generated by integrating the sub-slabs from different parts of the three-bulk a-Sb models. The lattice variation of c-Sb models upon downscaling is shown in Fig. 1c, showing that the in-plane lattice parameter *a* was gradually reduced from 0.436 nm (bulk) to 0.412 nm. This reduction in lattice parameter in ultrathin c-Sb films is consistent with the observation of the blue-shift in Raman spectroscopy experiments[35]. This lattice contraction phenomenon was also reported for other c-PCMs[50-52].

Prior to the electronic structure and optical calculations, all the bulk and slab models were further relaxed at 0 K. The semiconducting nature of 1 BL model is reproduced with our DFT calculation, and the calculated band structure is shown in Fig. 1c inset (gap size ~1.26 eV). For thicker models, we calculated their density of states (DOS) to enable a direct comparison between the crystalline phase (Fig. 1d) and the amorphous phase (Fig. 1e). A sizable energy gap can still be observed in the crystalline model of 2BL and the amorphous models of up to 4BL (see the data for two other amorphous configurations in Fig. S2), but for thicker models, the gap is filled.

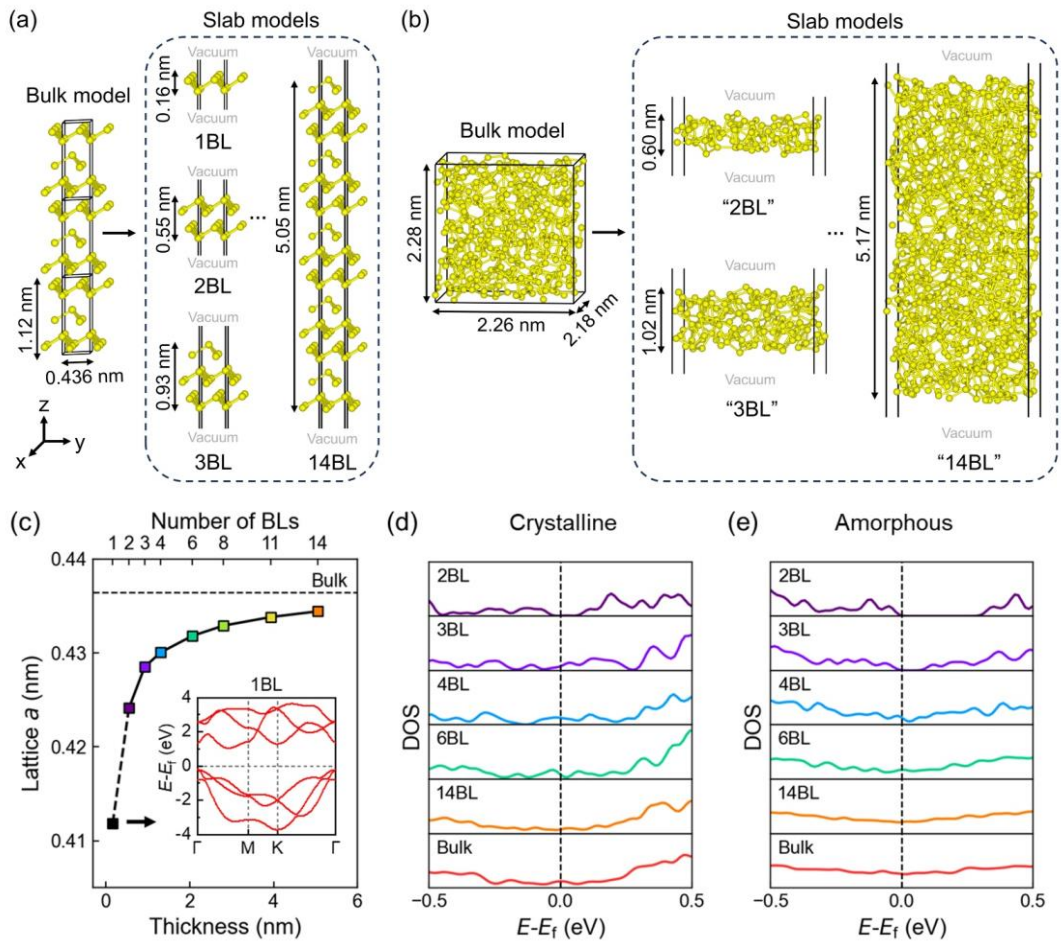

**Fig. 1. DFT modeling of bulk and slab models of Sb.** (a) The relaxed crystalline bulk and slab models. All crystalline models were built using hexagonal unit cells, except the z axis. (b) The calculated amorphous bulk and slab models. All amorphous models were built using orthorhombic cells. (c) The optimized in-plane lattice parameter of crystalline models as a function of thickness. The inset shows the band structure of 1 BL model. The DOS profiles calculated for (d) crystalline and (e) amorphous models.

Next, we carried out calculations of the real ($\varepsilon_1$) and imaginary ($\varepsilon_2$) parts of the dielectric function for these bulk and slab models. The frequency-dependent dielectric functions were calculated within the independent-particle approximation, which was shown to be adequate to characterize the optical properties of PCMs[53-57]. The refractive index ($n$) and extinction coefficient ($k$) were calculated using the following formulas[58]:

$$n = \left(\frac{\sqrt{\varepsilon_1^2 + \varepsilon_2^2} + \varepsilon_1}{2}\right)^{\frac{1}{2}}, \quad (1)$$

$$k = \left(\frac{\sqrt{\varepsilon_1^2 + \varepsilon_2^2} - \varepsilon_1}{2}\right)^{\frac{1}{2}}. \quad (2)$$

These frequency-dependent optical data can be used to make a direct comparison with experimental results.

The optical profiles of the as-deposited and post-annealed Sb thin films in the absence of capping layers were measured via spectroscopic ellipsometry experiments in Ref.[35]. According to their Raman spectroscopy measurements, no oxidation issue was found for the Sb thin films. Here, we re-plot the optical data measured for the Sb thin films of ~3–12 nm in Fig. 2a. As the thickness decreases, the refractive index $n$ of crystalline films shows an increase in the visible-light range, but a reduction in the near infrared range. A crossover is observed at ~1000 nm. A similar thickness-dependent trend in $n$ is observed in the amorphous films, but with smaller variations and an earlier onset of crossover at ~800 nm. As regards the extinction coefficient $k$, both crystalline and amorphous Sb films show a clear reduction as the thickness decreases, covering the visible-light and near infrared ranges. Also, a larger contrast window between the thinnest and thickest films is observed in the crystalline films than the amorphous ones.

These optical features are overall consistent with our DFT calculations, as shown in Fig. 2b. The dependence of $n$ with thickness in both crystalline and amorphous films show opposite trends in two different wavelength regions, with crossover points at ~1300 nm and ~1050 nm respectively. The same holds for the experimental data (Fig. 2a), although the crossover points occur at smaller wavelengths. As for $k$, the crystalline and amorphous films exhibit overall reduction with decreasing thickness. In addition, the thickness dependence of these properties is less pronounced in amorphous films than in crystalline films, implying a diminution of the optical contrast with the decrease in thickness. The data for the other two sets of amorphous models is consistent with Fig. 2b and is shown in Fig. S3. It should be noted that these DFT calculations were carried out with standard PBE functional, which is less accurate in describing the size of energy gap. Therefore, it is not feasible to obtain exactly the same optical values as experimental measurements. In addition, our DFT calculations are based on simplified models that cannot fully capture the polycrystalline nature of the experimental samples. Nevertheless, the qualitative trend in optical properties as a function of film thickness is well reproduced by our DFT calculations. In short, both experiments and DFT calculations consistently find a reduced optical contrast between a-Sb and c-Sb in the infrared spectrum as the film thickness decreases.

Moreover, our DFT calculations provide additional information to supplement experimental results. Our slab models cover the thickness range from 1 bilayer to ~5 nm, greatly extending the lower limit of experimental data (i.e., 3 nm). As shown in Fig. 2b, the optical properties of ultra-thin films (< 2 nm) are notably different from thicker films. In the visible-light region, both the refractive index $n$ and the extinction coefficient $k$ of crystalline ultra-thin films are visibly higher than for thicker films. In the near-infrared, a much lower $k$ is expected in either phase for ultra-thin films.

With the thicknesses-dependent optical data, we were able to perform a series of FEM simulations using COMSOL (see Methods) to investigate the contrast window of Sb-based photonic memory devices to approach the thickness limit. We focused on a standard silicon-on-insulator waveguide memory device with the height and width of the waveguide $h_{wg}$ and $w_{wg}$ being set as 0.12 μm and 0.45 μm, consistent with Ref. [36]. Regarding the parameters of the Sb film, we considered a normal material length $d_{Sb}$ = 4 μm, and varied the film thickness $h_{Sb}$ from 0.6 nm up to 12 nm. Based on the input and output power of waveguides ($P_1$ and $P_2$, respectively), we can calculate their transmittance ($T=P_2/P_1$) and optical loss ($-10\lg(P_2/P_1)$).

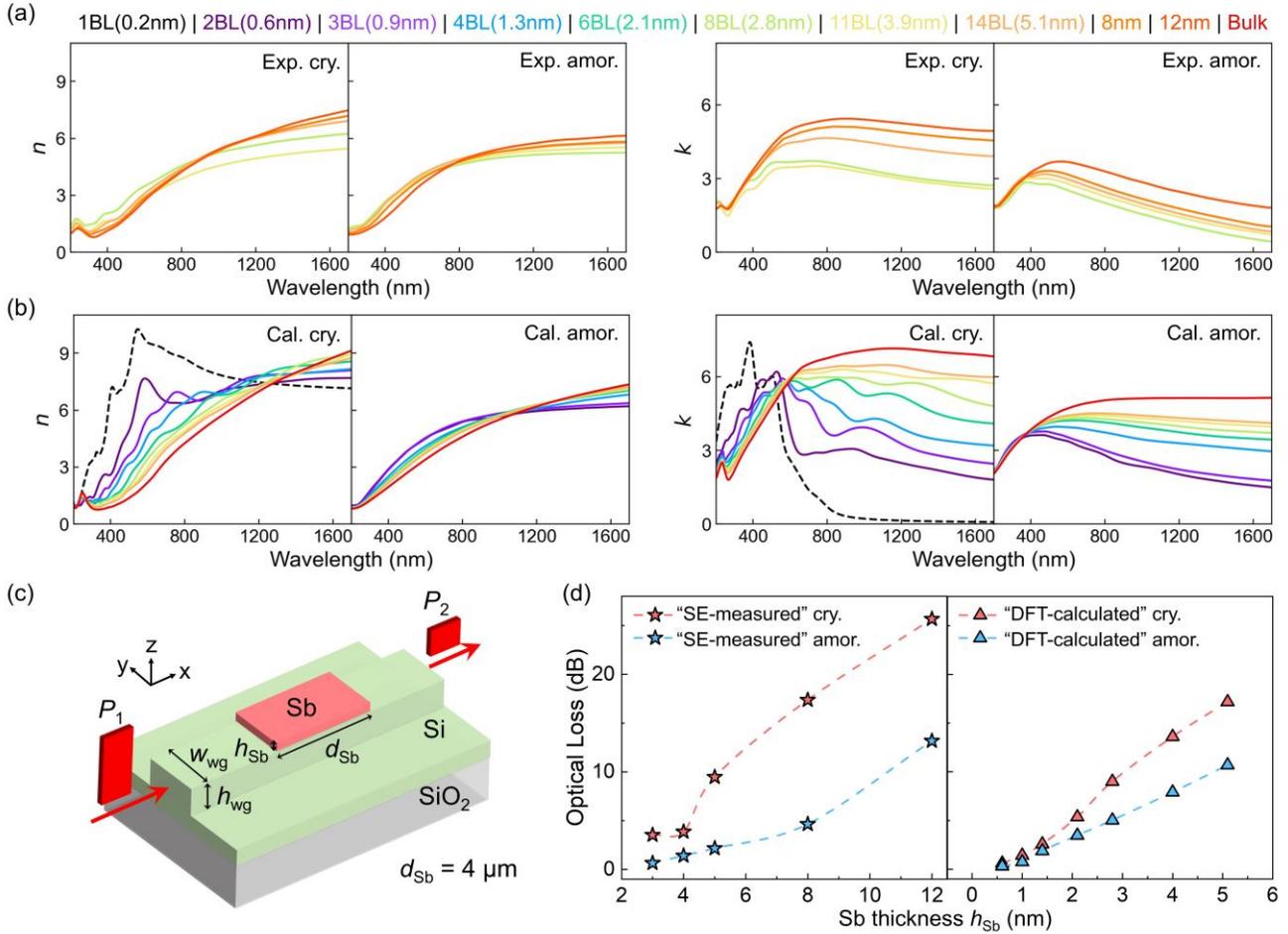

**Fig. 2. Comparison between optical properties of Sb from experiments and DFT calculations.** Refractive index ($n$) and extinction coefficient ($k$) in the spectrum range between 200 nm and 1700 nm for crystalline (cry.) and amorphous (amor.) antimony thin films with varying thicknesses. (a) The experimental (Exp.) data obtained by spectroscopic ellipsometry measurements, taken from Ref.[35] with permission. (b) The computational results (Cal.) have been obtained in this work. (c) A schematic of the antimony-based photonic waveguide studied in our FEM simulations. (d) The optical loss of waveguides at the telecommunication C-band (1550 nm) from FEM simulations. The optical data of Sb films from DFT calculations ("DFT-calculated", this work) and spectroscopic ellipsometry measurements ("SE-measured", Ref.[35]) were used to describe the optical response of antimony.

The calculated optical loss data as a function of $h_{Sb}$ are presented in Fig. 2d. For a given thickness $h_{Sb}$, we used both the measured and calculated $n$ and $k$ of thin film samples or models when available. We referred these two sets of FEM simulations as "spectroscopic ellipsometry (SE)-measured" and "DFT-calculated" in the following. A more pronounced reduction in optical loss was observed for the crystalline state than the amorphous state with the decrease of $h_{Sb}$, reducing the contrast window. This trend holds for both SE-measured and DFT-calculated data. For the amorphous phase, the change in optical loss follows a nearly linear trend as $h_{Sb}$ varies, while a sharper change was observed for the crystalline state in the range between ~5 nm and ~3 nm in the SE-measured curve and between ~3 nm (8 BL) and ~1.3 nm (4 BL) in the DFT-calculated curve. Notably, the contrast between the two phases got no longer visible at 1.3 nm (4 BL) and below.

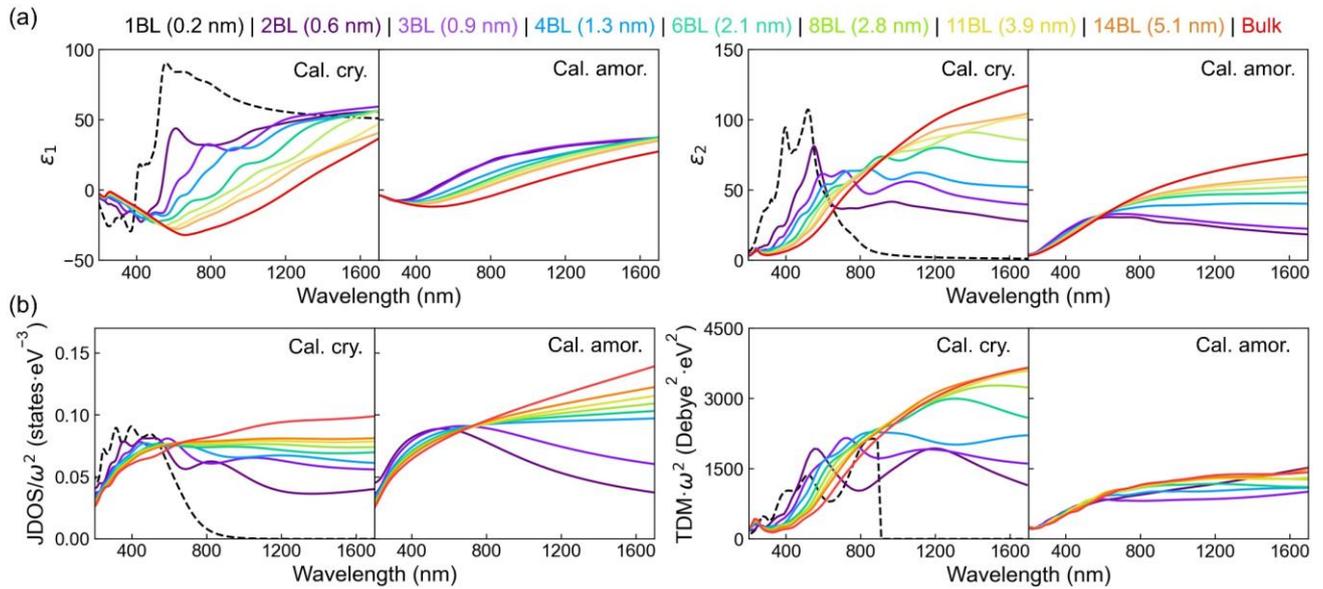

**Fig. 3. The real ($\varepsilon_1$) and imaginary ($\varepsilon_2$) parts of the dielectric function and the two contributing factors to $\varepsilon_2$.** (a) The real ($\varepsilon_1$) and imaginary ($\varepsilon_2$) parts of the frequency-dependent dielectric function for crystalline and amorphous antimony models. (b) Joint density of states (JDOS) and transition dipole moment (TDM). $\omega$ refers to the photon energy. The original TDM data contains a massive number of scatter points, with each point showing the transition probability of one possible excitation. Here the TDM points at each wavelength were averaged to make a line plot, for clarity.

We examine the underlying mechanism for such thickness-dependence of the optical properties. As revealed by formulas (1) and (2), the change in refractive index is a direct result of the evolution of the dielectric functions. Fig. 3a presents the calculated real ($\varepsilon_1$) and imaginary ($\varepsilon_2$) parts of the dielectric function. With the reduction of thickness, $\varepsilon_1$ increases in both crystalline and amorphous films. For $\varepsilon_2$, there is notable decrease in the infrared region for both states, and an additional increase below ~600 nm for the crystalline films. Importantly, crystalline films show more evident variations with thickness than amorphous films. To further explain such trend, the two contributing factors to $\varepsilon_2$, namely the joint density of states (JDOS) and the transition dipole moment (TDM)[53, 54], were calculated and presented in Fig. 3b. They quantify the amount of inter-band excitations and transition probability respectively, with the former one reflecting the effects of the electronic DOS and the latter one characterizing the degree of electron delocalization. For JDOS, both crystalline and amorphous phases show notable reduction with decreasing thickness in the near-infrared region, as a consequence of the opening of an energy gap and the resulting reduction in the DOS near the Fermi level. However, for TDM, crystalline films exhibit a dramatic reduction, while amorphous films show a much smaller variation. This observation implies that the decrease of film thickness results in localization of electrons in the crystalline phase but has modest impact on the amorphous phase, which accounts for the more obvious change in the optical properties in the crystal.

The evolution in the degree of electron delocalization can be further understood in terms of bonding mechanisms. It has been reported that amorphous PCMs exhibit ordinary covalent bonding, while crystalline PCMs are stabilized by metavalent bonding[59-67] (MVB). Distinct from ordinary covalent bonding based on electron pairing (i.e., two electrons per bond), MVB is characterized by extensive *p* orbital alignment and approximately one *p* electron per bond on average, giving rise to electron delocalization. As a result, metavalently bonded solids show a large chemical bond polarizability, leading to high values of the Born effective charge as well as a large optical dielectric constant[60]. But the *p* orbital alignment is broken in the amorphous phase, and the electrons become more localized, leading to a much weaker optical response. This is the reason why PCMs have a large optical contrast between the two states[68]. In previous works, two bonding indicators have been employed to distinguish MVB from other bonding types, namely the number of electrons transferred (ET) and electrons shared (ES) between pairs of neighboring atoms[62]. These quantities can be obtained from DFT calculations. MVB features intermediate ES values, i.e., smaller than those of ordinary covalent bonding but larger than metallic bonding values, as well as moderately small ET[62].

To analyze the evolution of the bonding character as a function of film thickness, we calculated the ET and ES values of the c-Sb models (see Methods), and presented them on an ET-ES bonding map (Fig. 4a). Since there is only one element in all the Sb models, their ET values are always zero. The ES values refer to the shortest bond of the particular atom. The ES value of a typical covalently bonded monatomic crystal (e.g. black phosphorus) is ~2, and that of a typical metallic crystal (e.g. silver) is ~0.5[68]. The ES value of Sb atoms in bulk rhombohedral model is thus calculated to be 1.47. For slab models, we calculated the ES values of each atom along the z axis, and the distributions are shown in Fig. 4b. For the 1BL model, the ES value is almost 2, indicating a highly localized bonding nature. As the thickness is doubled, the ES values of all four Sb atoms reduce sharply to 1.76. For thicker slab models, the bonding character of the atoms in the center shows a major difference as compared to the outer bilayers. The ES value of the former is gradually reduced from 1.56 to 1.48, but that of the atoms at the edge remains high, ~1.72. The ES value of the central atom of each slab model is plotted in Fig. 4a. As the thickness increases, the ES value changes from the position comparable to black phosphorus to that of bulk Sb, indicating a change in bonding character from covalent-like to metavalent-like. The 6BL (~2 nm) model seems to be a "critical" point, which also shows a large increase in TDM (Fig. 3b). These findings indicate that electron delocalization requires relatively long-range and collective orbital interaction and alignment.

Peierls distortion (PD), namely, the ratio of long and short bonds ($r_{long}/r_{short}$) in a well aligned bonding pair (Fig. 4c), is another important indicator in characterizing the bonding characteristics[66, 69, 70], which can be easily extended to highly disordered solids. We performed a layer-by-layer analysis of $r_{long}/r_{short}$ for both c-Sb and a-Sb models (Fig. 4d). The PD profiles of the c-Sb slab models provide insights consistent with the calculated ES values. Few-layer c-Sb slab models below 3BL show much more severe PD than the bulk c-Sb model. For thicker c-Sb slab models, the PD in the interior part always shows a lower value than that in the surface bilayers (the long Sb–Sb bonds are slightly increased by ~0.04 Å at the surfaces due to the less compact bonding environment), and is weakened further as an increase of thickness. We divided the amorphous slab models into layers according to the thickness of a single layer in bulk c-Sb (~0.2 nm), and calculated the average $r_{long}/r_{short}$ for each layer (Fig. 4d, right panel). Similar to the crystalline models, the surface layers (~0.7 nm thick) show larger PD than the center region. However, contrary to the crystalline models, where the weakening of PD with the increasing thickness is evident until 11BL (~4 nm), the amorphous films show bulk-like PD in the center as early as 4BL (~1.3 nm). Due to the disordered atomic structure of a-Sb, electron orbitals are misaligned, leading to localized orbital interactions. Therefore, the bonding properties of a-Sb are much less affected by the thickness, and the evolution in TDM is thus modest.

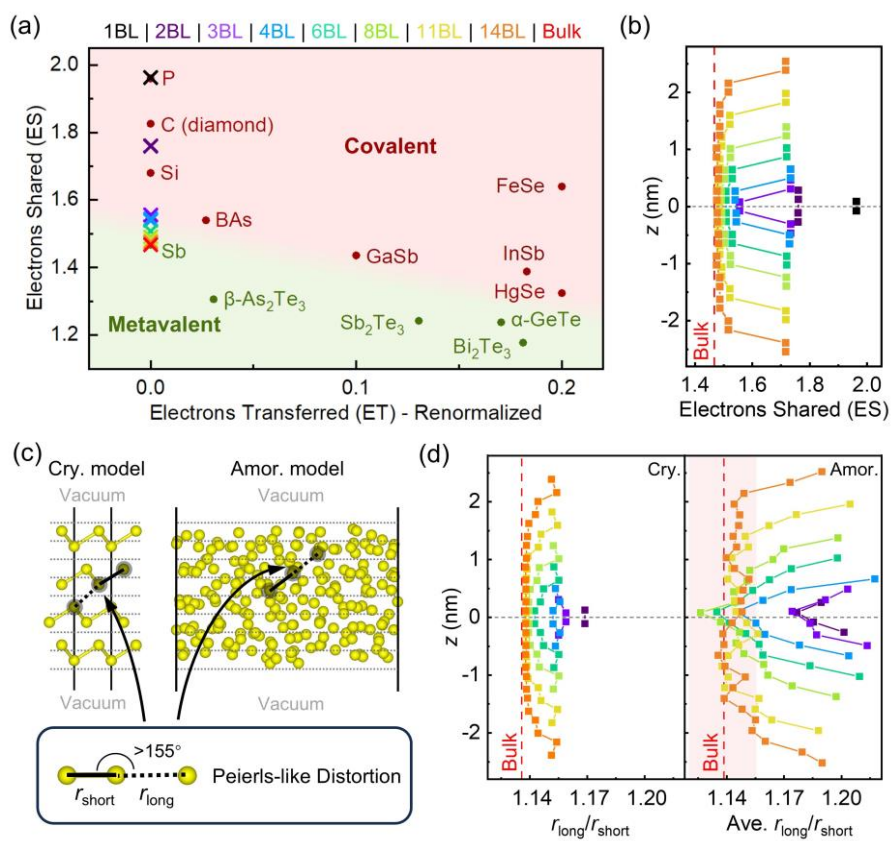

**Fig. 4. Chemical bonding analysis for crystalline and amorphous models.** (a) The locations of the crystalline models on the bonding map defined by the Electrons Transferred (ET) and Electrons Shared (ES) indicators. The crosses represent the antimony models in this work, while dots stand for data for other materials adapted with permission from Ref.[64]. (b) The ES for each atomic layer in crystalline slab models. $z$ refers to the coordinates of the layers along the $z$ axis. (c) A sketch illustrating the definition of Peierls distortion (PD). The grey dotted lines indicate the separation of each slab model into layers for PD analysis. (d) PD ($r_{long}/r_{short}$) for each layer in crystalline (Cry.) and amorphous (Amor.) models. The red shaded area shows the fluctuation range of PD in bulk amorphous models.

According to the analyses of PD (Fig. 4d) and coordination number (Fig. S4), the "4BL" and even thinner amorphous models show high structural dissimilarity as compared to the bulk amorphous phase. This feature indicates that nucleation could be difficult at or below such thickness of ~1.3 nm. Besides, a major change of bonding character from covalent-like to metavalent-like occurs in the range of 4–6 BL (1.3–2.1 nm) in the crystalline state, accompanied by an evident increase in excitation probability and hence a notable rise in optical contrast. Therefore, this range should be regarded as the bottom limit for a functional Sb thin film with both crystallization capacity and sizable contrast window.

To validate the theoretical prediction, we prepared Sb thin films via magnetron sputtering on silicon substrates with thicknesses of 1–7 nm. The thin films were all covered with a standard non-conductive capping layer of ~10 nm $ZnS:SiO_2$ for the electrical measurements as a function of time or temperature. As shown in Fig. 5a, the 7 nm a-Sb thin film started to crystallize after 100 seconds, but other thinner films can sustain the amorphous form over longer time. Upon in situ heating to 350 °C with a heating rate of 10 °C·min$^{-1}$, the 6 nm and 5 nm a-Sb thin films got crystallized quickly at $T_x$ of 58 and 77 °C, showing relatively poor thermal stability (Fig. 5b). Further downscaling largely enhanced the $T_x$ to be 145, 209 and 243 °C for the 4 nm, 3 nm, and 2 nm case, respectively. Note that the $T_x$ of a flagship PCM $Ge_2Sb_2Te_5$ is ~150 °C. Our Raman spectra (Fig. 5c) confirmed that after heating, these thin films indeed crystallized, as seen by the emergence of $E_g$ (at ~120 cm$^{-1}$) and $A_{1g}$ peaks (at ~155 cm$^{-1}$). No oxidation was observed in our samples, as the peak corresponding to $Sb_2O_3$ was absent in the Raman spectra. For even thinner films below 2 nm, however, the electrical resistance became too high, exceeding the measurable range, and there was no clear signature of crystallization from the Raman spectra.

To assess the thickness variation and uniformity, we carried out atomic force microscopy (AFM) experiment on an as-deposited a-Sb thin film of 2 nm with no capping layer. As shown in Fig. 5d, the overall thickness of the thin film was measured as ~1.97±0.16 nm, confirming an atomically flat surface morphology. However, direct thermal annealing to 350 °C would result in a cluster issue and the deposition of a capping layer is necessary to obtain 2 nm crystalline Sb thin films of high smoothness. Remarkably, this 2 nm Sb thin film still exhibited a resistance window of over four orders of magnitude (Fig. 5b), and the amorphous stability was largely enhanced with a high $T_x$ of 243 °C. These features make 2 nm Sb a superior option for electronic nonvolatile memory application as compared to thicker Sb films.

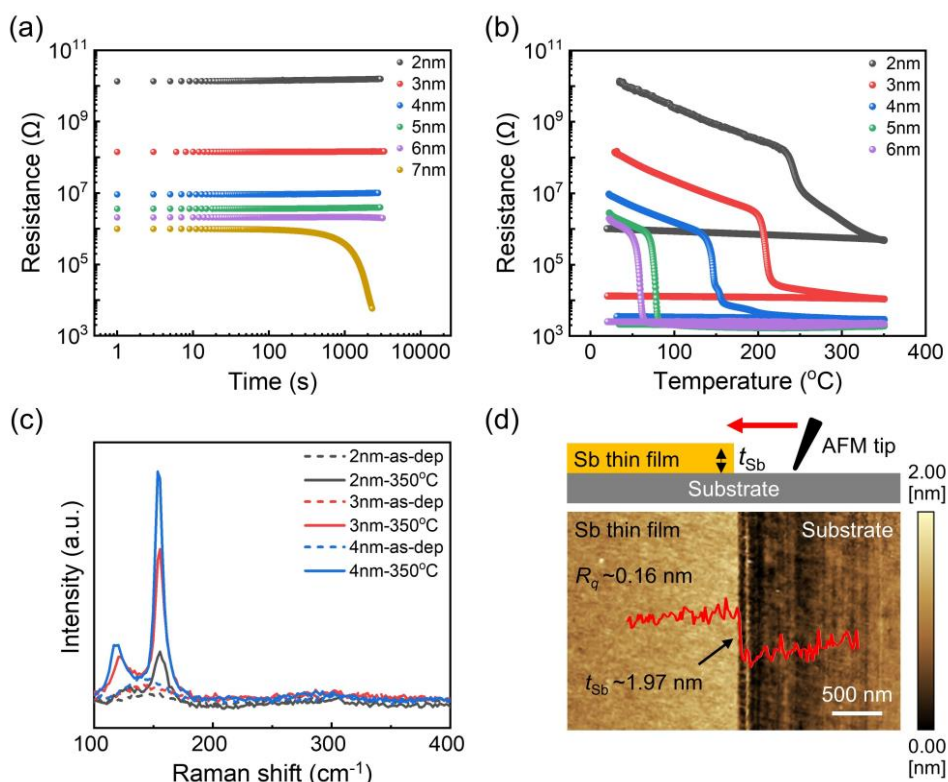

**Fig. 5. Preparation and characterizations of Sb thin films.** (a) The changes in sheet resistance of Sb thin films at room temperature over time. (b) The measured sheet resistance of Sb thin films as a function of temperature. (c) The Raman spectra of as-deposited and thermally annealed Sb thin films. (d) The atomic force microscopy measurements of ~2 nm as-deposited a-Sb thin film.

It is also noted that the overall electrical window was shifted towards higher resistance, indicating smaller optical absorption due to the reduction of free carriers. To assess the optical contrast, we carried out spectroscopic ellipsometry measurements on the as-deposited a-Sb thin films and the 350 °C thermally annealed c-Sb thin films, and all the Sb thin films were covered with a ZnS:SiO$_2$ capping layer. The raw ellipsometry data and the fitting parameters can be found in Fig. S6 and Table S1. As shown in Fig. 6a and 6b, the measured *n* and *k* values of 3 nm Sb thin films showed a clear contrast window induced by crystallization, and the contrast was consistently observed throughout the spectrum range of 400–1700 nm, with the average $\Delta n$ being ~0.57 and $\Delta k$ being ~0.87. As the thickness of Sb was reduced to 2 nm, the contrast window in *n* remained basically the same, and a moderately smaller contrast was found in *k* with $\Delta k$ being ~0.58. We note that different substrate and capping layer could affect the fitting of *n* and *k* data. In the above experiments, we used a standard silicon substrate that was covered by a native $SiO_2$ layer. Such substrate is frequently used for PCM-based optical applications. The capping layer $ZnS:SiO_2$ is non-conductive, transparent and colorless, and is frequently used as the protection layer for PCM. For comparison, we prepared as-deposited ~2 nm and ~3 nm Sb thin films on the same Si substrate but without capping layer. As shown in Fig. S7, the measured *n* and *k* data of the uncapped samples are comparable to those with capping layer (Fig. 6a and 6b). However, these *n* and *k* data are in general smaller than those reported by Cheng *et al.*[35]. This deviation may be due to the difference in the ellipsometry setup or the fitting approach. Nevertheless, the trend on reduced optical window in thinner films holds the same for both our measurements and those reported by Cheng *et al*.

This experimentally observed thickness-dependent optical trend was also consistent with our DFT calculations in Fig. 2b. To make a direct comparison, we extracted the data of the 6 BL (2.1 nm) and 8 BL (2.8 nm) models and re-plotted them in Fig. S5, in which the 8BL model also showed a larger $\Delta k$ than the 6BL model. Therefore, we conclude that this optical trend with thin thickness is an intrinsic volume effect, which stems from the bonding differences in the crystalline phase and the amorphous phase with and without *p* orbital alignment. The sizable optical contrast of 2 nm Sb in the range of 400–1700 nm, makes it promising to develop non-volatile color display and all-optical photonic memory devices with enhanced state retention based on such two-dimensional Sb.

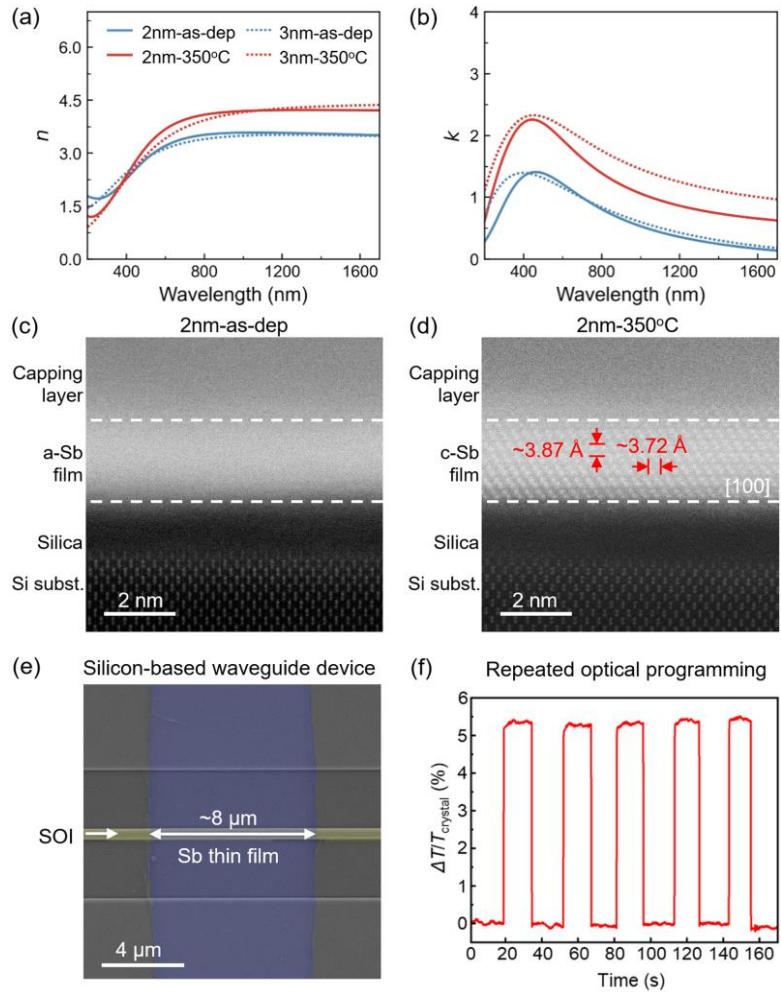

**Fig. 6. Phase-change induced optical contrast in Sb thin films approaching the two-dimensional limit.** (a) and (b) The refractive index *n* and extinction coefficient *k* of ~2 nm and ~3 nm Sb thin-films measured by spectroscopic ellipsometry. (c) and (d) The HAADF images of the as-deposited and thermally annealed Sb thin films of ~2 nm. (e) SEM image of an SOI waveguide device (yellow false color) decorated with a 2-nm-thick Sb thin film of length ~8 μm (cyan false color). (f) Repeated all-optical programming of the 2-nm-thick Sb waveguide device.

To gain an in-depth understanding of the atomic structures on the 2 nm Sb thin films, we performed cross-sectional STEM experiments in the High Angle Annular Dark Field (HAADF) mode. The atomic-scale images of the as-deposited a-Sb and thermally annealed c-Sb are presented in Fig. 6c and 6d, respectively. The former shows no visible lattice spots, and the latter shows clear Sb bilayer patterns with approximately 6 intact bilayers with a [100] zone axis. The measured in-plane and out-of-plane atomic spacings of 2 nm c-Sb are 3.72 and 3.87 Å, as marked by the red arrows. Therefore, we fully demonstrated that the Sb thin film can still undergo phase transition at a very limited thickness of 2 nm, corresponding to only 12 atomic layers in the crystalline phase. We note that such thickness limit in Sb is comparable to that of GeTe thin films obtained by the step-by-step growth via molecular beam epitaxy, in which 6 BL was also the bottom thickness limit to obtain crystalline GeTe[49]. We note that the crystallization of ultrathin Sb films (3–4 nm) with a well-ordered out-of-plane crystal orientation was also observed in Ref. [71], where no capping layer was deposited. But our AFM experiments showed that amorphous Sb formed a discontinuous crystalline thin film (~3 nm thick) with a very rough surface morphology after thermal annealing (Fig. S8). The measured root-mean-square roughness $R_q$ value of the annealed sample ~10.122 nm is much larger than that of the as-deposited amorphous sample ~0.159 nm. The same trend holds for the thinner Sb thin film of ~2 nm. Therefore, capping layers are necessary to avoid this clustering issue in our case. This discrepancy could be attributed to the different preparation method and thermal condition, i.e., pulsed laser deposition with a deposition temperature of 210 °C (Ref. [71]) and magnetron sputtering with subsequent thermal annealing at 300 °C and above (our work), which could lead to different crystallization pathways of ultrathin Sb films.

To prove that the 2 nm Sb thin film can still be reversibly switched optically, we fabricated silicon-on-insulator (SOI) waveguides and carried out all-optical pump-probe experiments. The back-end-of-line (BEOL) fabrication processes were performed via the multi-project wafer (MPW) service provided by the Chongqing United Microelectronics Center Co., Ltd. (CUMEC) using the 180-nm Complementary Metal Oxide Semiconductor (CMOS) node technology. Prior to the deposition of Sb thin film, it is important to perform reverse sputtering using argon plasma. Then we deposited a Sb thin film with ~2 nm in thickness and ~8 μm in length. More fabrication details can be found in Methods. Figure 6e shows a scanning electron microscopic (SEM) image of the fabricated Sb waveguide device. The silicon waveguide core and Sb thin film are highlighted by the yellow and cyan false colors, respectively. Before optical testing, the device was annealed to obtain a fully crystalline state to allow better programming consistency. We next performed reversible all-optical switching of the Sb waveguide devices using femtoseconds (fs) laser pulses[66, 69, 70]. Specifically, we applied 4 high-power laser pulses and 200 weak laser pulses with an identical pulse width of 800 fs to program the Sb waveguide device into an ON state with higher transmittance ($T_{amor}$) and an OFF state with lower transmittance ($T_{cryst}$), respectively. Figure 6f shows the measured transmittance response of the Sb waveguide device upon programming. A clear switching contrast $\Delta T/T_{cry}$ = ~5.3%, where $\Delta T = T_{amor} - T_{cryst}$, was observed. The repeated optical programming demonstrated that ~2 nm Sb can act as a functional PCM for the photonic waveguide memory. We also note that Aggarwal *et al.* reported reversible switching of ~3 nm Sb waveguide devices[66, 69, 70], but the transmittance response of their devices was less stable over time, i.e., after sending amorphization laser pulses, the change in the optical readout firstly reached 12%, but got reduced to 2–6% in a few seconds. By reducing the film thickness further to ~2 nm, the much-improved amorphous stability can lead to more robust device performances. Nevertheless, the fabrication and optical testing of Sb-based devices with such thickness get more challenging, and more efforts are needed to evaluate device variabilities in high-density arrays.

At last, we investigated the properties of closely related single-element materials, namely, tellurium[72] and bismuth[73], and evaluate whether they can be used as monatomic PCM. The DFT calculations showed the ES value of crystalline Te and Bi to be 1.68 and 1.41, respectively. As shown in the ET-ES bonding map in Fig. S9, Te falls in the covalent bonding region, while Bi lies in the metavalent bonding region. This calculation suggests that Te may not be a suitable PCM. Indeed, according to Zhu *et al.*[72], Te behaves as a monatomic phase-change switch (PCS) material[74-76] rather than a PCM. More specifically, crystalline Te (with a sizable bandgap) served as the high-resistance OFF state, and liquid Te (with metallic character) as the low-resistance ON state. This is a volatile phase transition, since the 20-nm-thick Te film crystallizes spontaneously after programming[72]. In principle, it should be possible to obtain amorphous Te by reducing the film thickness to ~2–3 nm, similar to the Sb case.

We have prepared ~2 nm and ~3 nm as-deposited Te thin films and performed TEM and electrical measurements on them. As shown in Fig. S10a, a blurred halo ring is observed in the selected-area electron diffraction (SAED) pattern of the ~2 nm Te thin film, whereas sharp diffraction spots appeared in that of the ~3 nm Te film, indicating that amorphous Te can only be stabilized when the film thickness is reduced to ~2 nm. During *in situ* heating to 200 °C, the room-temperature resistance first decreases but then increases. This can be attributed to the fact that crystalline Te also tends to form a high-resistance state. As a result, the resistance contrast is very limited and, thus, elemental Te cannot be considered a PCM, even though amorphous Te can be stabilized through confinement effects. In addition, the resistance values of the ~2 nm Te thin films exceed $10^{10}$ Ω, which are too high for practical PCM applications.

According to the bonding calculation, Bi could be a potential candidate for PCM applications. However, it turns out to be difficult to obtain amorphous Bi through confinement effects. The SAED patterns in Fig. S10b show that both the ~2 nm and ~3 nm Bi thin films were already crystallized upon deposition. The resistance values barely changed after *in situ* heating. Hence, even a thickness of ~2 nm is not sufficient to obtain as-deposited amorphous Bi thin films. In other words, the crystallization tendency of Bi is stronger than that of Sb. It might still be possible to obtain amorphous Bi in an even thinner film, e.g., around 1 nm. Nevertheless, such ultrathin films would be more challenging to use for practical applications. Overall, the unique balance between crystallization kinetics and property contrast makes Sb the only monatomic PCM successfully realized so far.

## 3. Conclusion

We have provided an atomistic understanding on the thickness-dependent optical and bonding properties of monatomic PCM antimony by performing systematic *ab initio* simulations and FEM simulations. Reduction in electronic density of states near the Fermi level was found to contribute to the decrease of $k$ in both crystalline and amorphous films. The weakening of MVB upon downscaling to a few atomic layers significantly impairs the optical excitation probability in the crystal, resulting in lower $\Delta k$ for thinner films. Nevertheless, the $\Delta k$ should be sufficiently large to determine the ON and OFF states for data encoding. Scaling the thickness of Sb thin film down to 3 nm and below, capping layers became necessary in preventing the film rupture issue. With joint theoretical and experiment efforts, we demonstrated 2 nm to be the bottom thickness limit for functional Sb thin films, enabling a more robust amorphous stability with $T_x$ ~243 °C for nonvolatile electronic applications. Considering the sizable changes in electrical resistance, extinction coefficient in the near infrared range and refractive indices in the visible light range, the 2 nm Sb thin film is suitable for nonvolatile photonic memory and reflective optical display applications. Our optical experiments demonstrate that the 2 nm Sb waveguide device can be reversibly programmed, exhibiting a sizable switching contrast of ~5.3% Our combined multiscale simulations[77] and experiments should serve as a typical example on how to optimize the thickness limit of materials for nonvolatile and reconfigurable optical applications from the atomic scale.

## 4. Methods

***Materials modeling***: *Ab initio* molecular dynamics simulations for the amorphous models were performed using the second-generation Car-Parrinello scheme[78] as implemented in the CP2K package[79], with identical computational setup as in Ref.[38]. We used the Perdew–Burke–Ernzerhof (PBE) functional[80] and Goedecker pseudopotentials[81] along with Grimme's D2 dispersion correction[82]. Only the $\Gamma$ point was used to sample the Brillouin zone, and the timestep was set to 2 fs. Structural relaxations and optical calculations were carried out using the Vienna Ab-initio Simulation Package (VASP)[83]. The PBE functional and projector augmented-wave (PAW) pseudopotentials[84] were used with an energy cutoff of 500 eV, and the Grimme's D3 scheme[85] was employed for dispersion corrections. Gaussian smearing with width of 0.05 eV was used to describe partial occupancies of orbitals. For structural relaxation, the *k*-point meshes were 13×13×5 and 13×13×1 respectively for crystalline bulk and slab models, and only the $\Gamma$ point was used for the amorphous models. The convergence criteria were set as $10^{-5}$ eV for electronic self-consistent loops and 0.01 eV/Å for ionic relaxation. For optical calculations, the *k*-point density along each lattice direction was increased to at least twice for all the models for better convergence of the calculated optical properties, with a more stringent convergence threshold of $10^{-6}$ eV. The fully relaxed structures were then used for self-consistent calculations using Quantum ESPRESSO[86], as the obtained electronic wavefunctions can be directly processed by the Critic2 code[87] to compute the bonding indicators[67, 68]. The QE calculations were applied with both norm-conserving[88] and PAW potentials together with PBE (the energy cutoff for wavefunctions and charge density was set as 80 and 320 Ry, respectively). In Critic2 calculations, the Bader's basins are determined based on wavefunctions, and then the Domain Overlap Matrices (DOM) are calculated to yield the delocalization (DIs) and localization indices (LIs), which measure the quantity of electron pairs shared between two basins and the number of electrons localized in a basin respectively. The electrons shared (ES) between each pair of atoms is the twice of DI, and the electron transfer (ET) of an atom is determined by subtracting the number of electrons for the free reference atom from the number of localized electrons in its basin, and by dividing the resulting figure by the formal oxidation state.

***COMSOL simulations***: We used the RF module of COMSOL Multiphysics to perform finite element method simulations for waveguides working at 1550 nm band. Frequency domain simulations were run after boundary mode analyses of input and output ports. The refractive indices of silica and silicon were described by the built-in data provided by the software. Perfectly matched layer (PML) boundary condition was applied to truncate the simulation region in three dimensions, and the distance between PML and silicon waveguide is around 2 μm. The meshes in each material region were constructed with the maximum mesh size being smaller than one fifth of the effective wavelength.

***Thin film preparation and characterizations:*** The Sb thin films were deposited onto $SiO_2$/Si substrate at room temperature via magnetron sputtering (AJA, Orion-8), using a Sb target (99.99%) at pressure of 4.7 mTorr with the radio frequency (RF) at ~10 W in the high vacuum sputtering chamber. The thickness of Sb was controlled by the sputtering time, and the deposition rate was ~0.72 nm·min$^{-1}$. A ZnS:$SiO_2$ capping layer of ~10 nm was deposited on top of Sb, except for the sample used in atomic force microscopy (AFM) experiments (using a SPM-9700HT). Electrical measurements were performed using a Keithley 2636B source meter and an Instec mK200 hot stage instrument. The resistance of the films was measured in situ in an Ar atmosphere as a function of temperature using a two-point probe method with a heating rate of 10 °C·min$^{-1}$, where the probe electrode was titanium. The Raman spectra were collected by using a Renishaw inVia Qontor Raman microscope with a solid-state 532 nm laser for excitation. The laser power was set to 0.25 mW, and the exposure time was 1 s with 50 cycles. The spectroscopic ellipsometry measurements were performed with a UVISEL PLUS ellipsometer. The incidence angle was set to 70° with the light source of xenon lamps. The refractive indices of Sb were obtained by fitting the measured spectra using the CODE software (www.mtheiss.com) based on a multi-layer model involving the substrate, the Sb thin film and the capping layer. Two roughness layers were integrated into the established system, and were positioned at the surface of the capping layer and at the interface between the capping layer and the Sb film, respectively. Both roughness layers were described using the Bruggeman effective medium approximation with a fixed volume fraction of 0.5. The dielectric model of amorphous Sb included a constant dielectric background and a Tauc–Lorentz oscillator, while an additional Drude contributor was incorporated for the crystalline phase. The dielectric functions of the substrate and the capping layer were determined independently based on reference samples (additional fitting details can be found in the caption of Table S1). Such method has been established in optical measurements of typical PCMs. The focused-ion beams facility (FIB, Hitachi NX5000) was used to fabricate cross-sectional TEM specimens. The high-angle annular dark field scanning transmission electron microscopy (HAADF-STEM) imaging experiments were performed on a JEM-ARM200F microscope with double spherical aberration correctors. The SEM experiments were made using a Zeiss Sigma 300 scanning electron microscope.

***Waveguide device fabrication:*** The shallow-etched SOI waveguides, featuring an etching depth of 150 nm and a width of 450 nm, were taped out via the MPW service provided by CUMEC. A selected region of the silicon dioxide upper cladding on the SOI waveguide was removed by reactive-ion etching (RIE) to enable the subsequent deposition of the Sb thin film. Reverse sputtering using argon plasma etching was then carried out at a radio frequency power of 10 W for 5 minutes. Photolithography was performed to define the PCM deposition window using a negative-tone photoresist (AZ 5214). A 2-nm-thick Sb thin film was deposited on the silicon waveguides via magnetron sputtering (AJA Orion-8) with a sputtering power of 10 W, a working pressure of 4.7 mTorr, and a deposition rate of ~0.72 nm·min$^{-1}$, calibrated from a prior Sb thin-film deposition used for structural and optical characterizations. The Sb layer was subsequently capped with a ~10-nm-thick indium–tin–oxide (ITO) layer. Finally, the photoresist was removed to lift off the Sb and ITO thin films outside the patterned windows, leaving patch-shaped Sb/ITO stacks integrated on the silicon waveguides.

## Supporting Information

Supporting Information is available from …

## Acknowledgements

We acknowledge useful discussions with Prof. Harish Bhaskaran and Prof. Zengguang Cheng. We thank Ruobing Lin for his help on the deposition and characterizations of Te thin films. The work is supported by the National Key Research and Development Program of China (2023YFB4404500). W.Z. thanks the support of the National Natural Science Foundation of China (62374131). The HPC platform of XJTU and the National Supercomputing Center in Xi'an are acknowledged for providing computational resources. The International Joint Laboratory for Micro/Nano Manufacturing and Measurement Technologies of XJTU is acknowledged. R.M. gratefully acknowledges funding from the PRIN 2020 project "Neuromorphic devices based on chalcogenide heterostructures" funded by the Italian Ministry for University and Research (MUR). M.W. acknowledges funding from Deutsche Forschungsgemeinschaft within SFB 917 "Nanoswitches".

## Conflict of Interest

The authors declare no conflict of interest.

**Data Availability Statement**

Data supporting this work will be available at https://caid.xjtu.edu.cn/info/1003/2071.htm upon journal publication.

# Supporting Information

Atomistic understanding of two-dimensional monatomic phase-change material for non-volatile optical applications

*Hanyi Zhang[#], Xueqi Xing[#], Jiang-Jing Wang\*, Chao Nie, Yuxin Du, Junying Zhang, Xueyang Shen, Wen Zhou\*, Matthias Wuttig, Riccardo Mazzarello\*, Wei Zhang\**

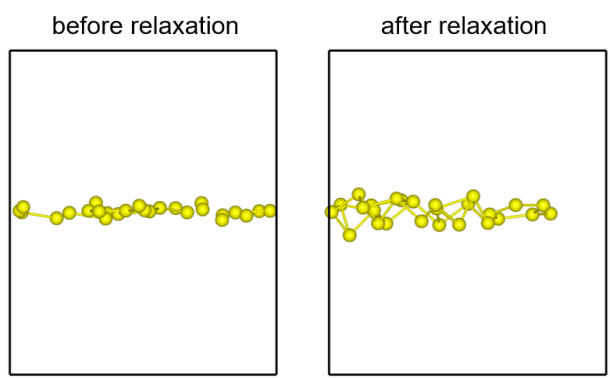

**Figure S1.** The atomic coordinates of the thinnest a-Sb model before and after relaxation. The initial coordinates were taken from bulk a-Sb model along the vertical direction over ~0.2 nm. The limited thickness leads to clustering of atoms and a rupture of the slab.

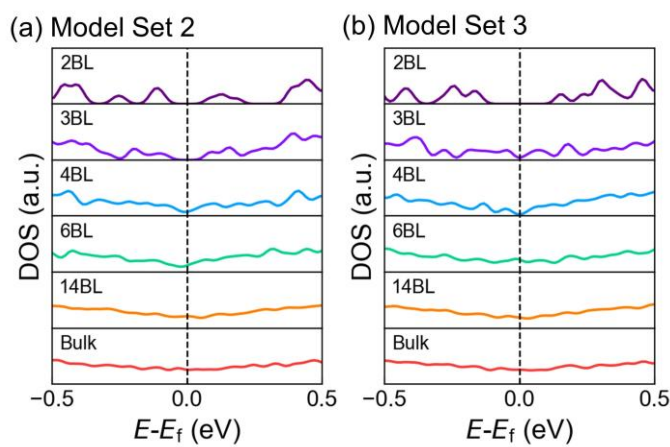

**Figure S2.** The electronic density of states (DOS) for (a) the second and (b) the third sets of amorphous models.

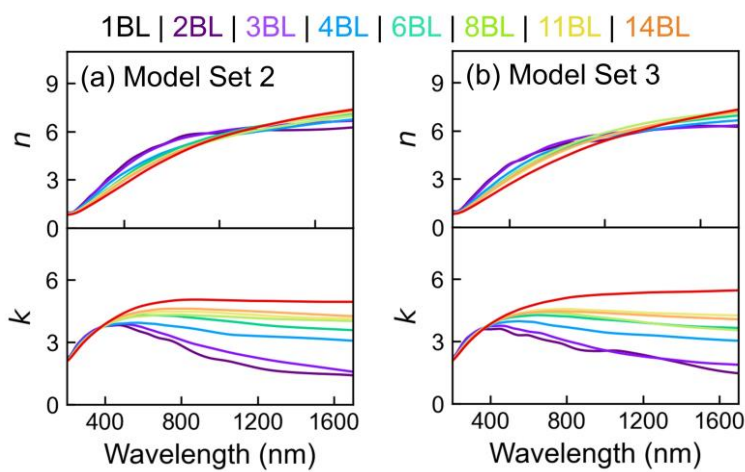

**Figure S3.** The refractive index ($n$) and extinction coefficient ($k$) for (a) the second and (b) the third sets of amorphous models.

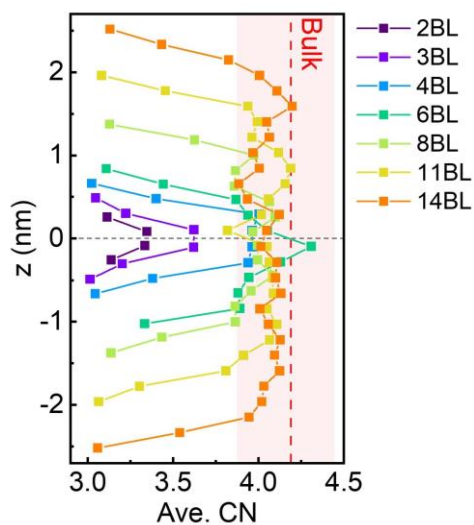

**Figure S4.** Average coordination number for each layer in amorphous models. The interatomic cutoff is set as 3.4 Å. The red shaded area shows the fluctuation range of coordination number in bulk amorphous models.

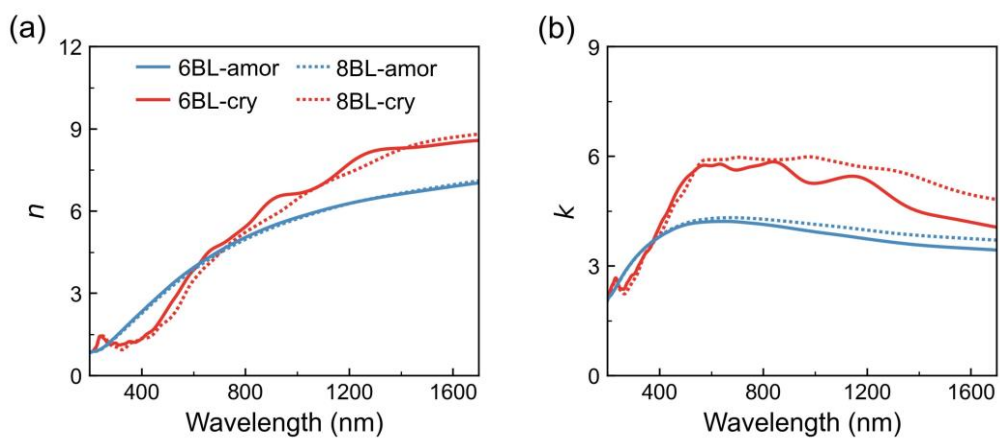

**Figure S5.** The DFT-calculated refractive index *n* and extinction coefficient *k* of the 6BL (2.1 nm) and 8BL (2.8 nm) Sb models, as extracted from Fig. 2b. Note that these calculations were performed using the PBE functional, which is known to underestimate the size of the energy gap, leading to stronger optical responses. Nevertheless, the thickness-dependent optical trend was consistently observed in both DFT calculations and ellipsometry experiments.

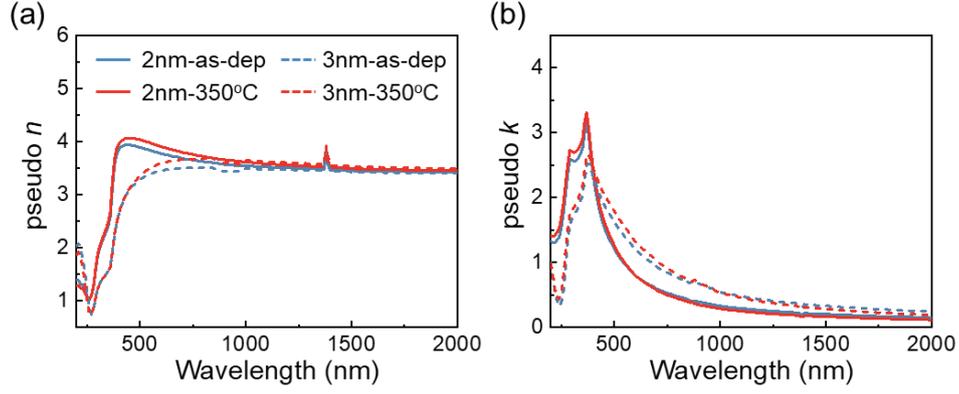

**Figure S6.** Raw ellipsometry data, namely, (a) pseudo *n* and (b) pseudo *k*, corresponding to the *n* and *k* data fitting of the Sb thin films in Fig. 6a and 6b.

|  | 2nm-as-dep | 2nm-350°C | 3nm-as-dep | 3nm-350°C |
|---|---|---|---|---|
| $\omega_0$ (eV) | 2.54 | 2.61 | 3.02 | 2.64 |
| $A$ (eV) | 33.21 | 43.44 | 46.02 | 50.61 |
| $\omega_\tau$ (eV) | 3.04 | 2.98 | 5.00 | 3.84 |
| $\omega_{\text{Gap}}$ (eV) | 0.40 | 0.13 | 0.39 | 0.06 |
| $\varepsilon_{\text{const}}$ | 4.57 | 3.69 | 3.29 | 2.35 |
| $\Omega_P$ (eV) | - | 1.68 | - | 1.96 |
| $\Omega_\tau$ (eV) | - | 2.42 | - | 2.40 |
| Sb thickness (nm) | 2.00 | 1.86 | 3.19 | 2.80 |
| Sb/capping roughness layer thickness (nm) | 0.30 | 0.10 | 0.28 | 0.42 |
| Capping thickness (nm) | 4.98 | 4.80 | 10.00 | 9.50 |
| Capping surface thickness (nm) | 0.10 | 0.11 | 0.11 | 0.16 |

**Table S1.** Final fitting parameters obtained from the CODE calculations for the ellipsometry analysis of 2 nm and 3 nm Sb thin films with capping layers. The dielectric model of amorphous Sb consisted of a constant background term, $\varepsilon_{\text{const}}$, and a Tauc–Lorentz oscillator. The Tauc–Lorentz model expresses the imaginary part of the susceptibility as:

$$\chi(\omega) = \frac{1}{\omega} \frac{A\omega_0\omega_\tau(\omega - \omega_{\text{Gap}})^2}{(\omega^2 - \omega_0^2)^2 + \omega^2\omega_\tau^2} \Theta(\omega - \omega_{\text{Gap}})$$

where $A$, $\omega_\tau$, $\omega_0$ and $\omega_{\text{Gap}}$ are fitting parameters representing the amplitude, damping constant, resonance frequency and optical bandgap, respectively. An additional Drude term was incorporated to describe the free-carrier contribution in the crystalline phase:

$$\chi(\tilde{\nu}) = -\frac{\Omega_p^2}{\tilde{\nu}^2 + i\tilde{\nu}\Omega_\tau}$$

where $\Omega_P$ and $\Omega_\tau$ are the fitting parameters for plasma frequency and damping constant.

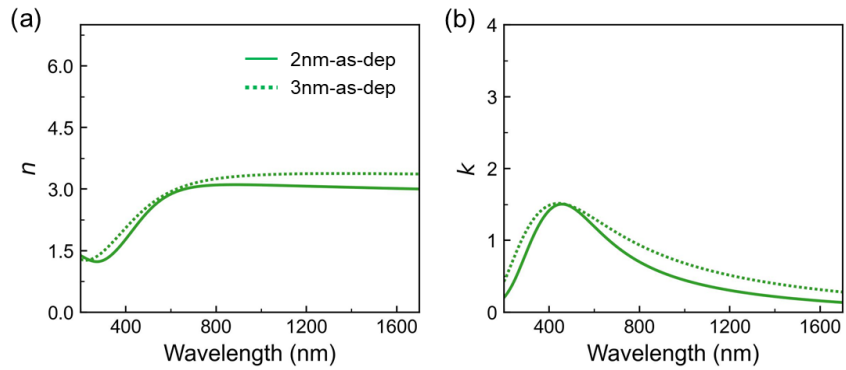

**Figure S7.** Measured refractive index $n$ and extinction coefficient $k$ of as-deposited ~2 nm and ~3 nm Sb films without capping layers.

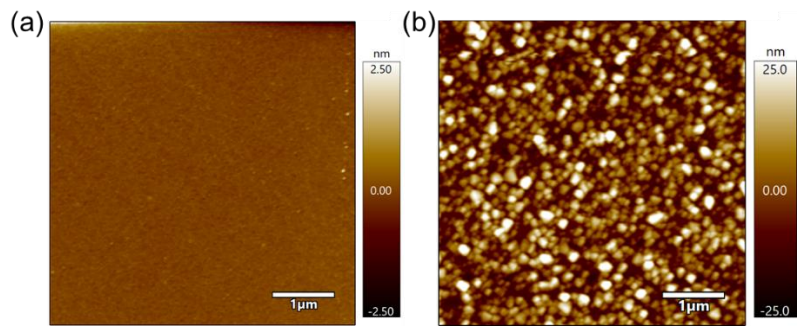

**Figure S8.** AFM images of ~3 nm Sb thin films without a capping layer. The two sets of samples were prepared in a single sputtering run at room temperature. (a) AFM image of the first as-deposited amorphous Sb thin film, showing a smooth surface with a root-mean-square roughness ($R_q$) of ~0.159 nm. (b) AFM image of the second Sb sample after annealing at 300 °C. The measured $R_q$ value of ~10.122 nm is much larger than that of the initial as-deposited amorphous sample.

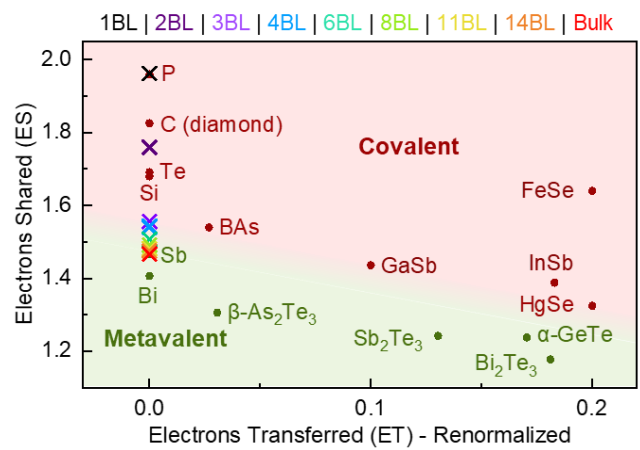

**Figure S9.** The ES data of crystalline Te (1.68) and Bi (1.41) are included in the bonding map.

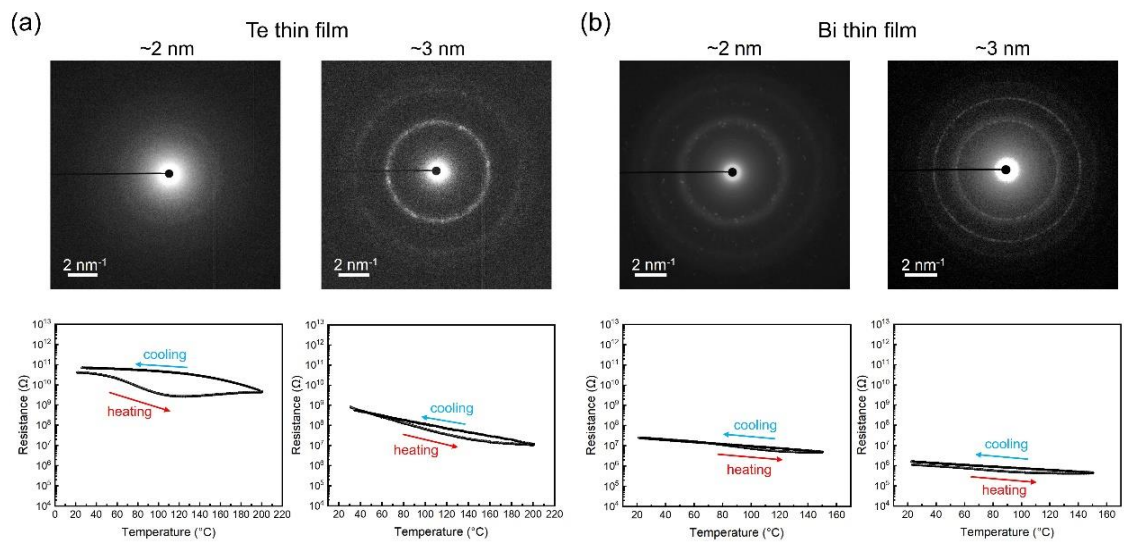

**Figure S10.** The SAED pattern and the resistance–temperature measurements of the as-deposited (a) Te and (b) Bi thin films with a thickness of ~2 nm and ~3 nm. The Te and Bi thin films were deposited onto the same silicon substrate at room temperature via magnetron sputtering, using a Te target (99.99%) and a Bi target (99.99%) at pressure of 4.7 mTorr with the radio frequency (RF) at ~10 W in the high vacuum sputtering chamber. A ZnS:SiO2 capping layer of ~10 nm was deposited on top of these thin films.